\documentclass[12pt]{article}
\usepackage{amssymb,amsmath,cite,epsfig}
\usepackage[top=2.5cm, bottom=2.5cm, left=2.5cm, right=2.5cm]{geometry}

\input{tcilatex}
\begin{document}

\begin{titlepage}
\vspace*{3cm}
\begin{center}
{\Large \textsf{\textbf{Exact solutions of the 2D Schr\"{o}dinger equation with central potentials induced by the non-commutativity of space}}}
\end{center}
\vskip 5mm
\begin{center}
{\large \textsf{Slimane Zaim$^{*}$}}\\
\vskip 5mm
D\'{e}partement des Sciences de la Mti\`{e}re, Facult\'{e} des Sciences,\\
Universit\'{e} Hadj Lakhdar -- Batna, Algeria.\\
\end{center}
\vskip 2mm
\begin{center}{\large\textsf{\textbf{Abstract}}}\end{center}
\begin{quote}
We obtain exact solutions of the 2D Schr\"{o}dinger equation with the central potentials
$V(r)=ar^2+br^{-2}+cr^{-4}$ and $V(r)=ar^{-1}+br^{-2}$ in a non-commutative space up to the first order of noncommutativity parametert  using the power-series expansion method
similar to the 2D Schr\"{o}dinger equation with the singular even-power and inverse-power potentials respectively in commutative space.
We derive the exact non-commutative energy levels and show that the energy is shifted to $m$ levels, as in the Zeeman effect.
\end{quote}
\vspace*{2mm}


\noindent\textbf{\sc Pacs numbers}:  02.40.Gh, 03.65.Ge, 03.65.Fd
\vskip 20mm
\vspace*{60mm}
{ \textsf{$^{*}$Corresponding Author, E-mail: zaim69slimane@yahoo.com}}\
\end{titlepage}

\section{Introduction}

Non-commutative quantum mechanics is motivated by the natural extension of
the usual quantum mechanical commutation relations between position and
momentum, by imposing further commutation relations between position
coordinates themselves. As in usual quantum mechanics, the non-commutativity
of position coordinates immediately implies a set of uncertainty relations
between position coordinates analogous to the Heisenberg uncertainty
relations between position and momentum; namely: 
\begin{equation}  \label{eq:1}
\left[ x^{\mu },x^{\nu }\right] _{\ast }=i\theta ^{\mu \nu }\,,
\end{equation}
where $\theta ^{\mu \nu }$ are the non-commutativity parameters of dimension
of area that signify the smallest area in space that can be probed in
principle. We use the symbol $\ast $ in equation \eqref{eq:1} to denote the
product of the non-commutative structure. This idea is similar to the
physical meaning of the Plank constant in the relation $\left[ x_{i},p_{j}%
\right] =i\hbar \delta _{ij}$, which as is known is the smallest phase-space
in quantum mechanics.

Our motivation is to study the effect of non-commutativity on the level of
quantum mechanics when space non-commutativity is accounted for. One can
study the physical consequences of this theory by making detailed analytical
estimates for measurable physical quantities and compare the results with
experimental data to find an upper bound on the $\theta $ parameter. The
most obvious natural phenomena to use in hunting for non-commutative effects
are simple quantum mechanics systems with central potential, such as the
hydrogen atom \cite{1,2,3,3a,3b}. In the non-commutative space one expects
the degeneracy of the initial spectral line to be lifted, thus one may say
that non-commutativity plays the role of magnetic field.

It has recently been shown that the non-inertial motion of the atom also
induces corrections to the Lamb shift \cite{4,5,6,7}. However, all the
aforementioned studies are concerned with flat space-time. Therefore, it
remains interesting to see what happens if the atom with central potential
is placed in a non-commutative space rather than a flat one. In this work we
present an important contribution to the non-commutative approach to the Schr%
\"{o}dinger equation with central potentials. The study of the exact and
approximate solutions of Schr\"{o}dinger equation with central potentials
has proved to be fruitful and many papers have been published $\left[ 1-42%
\right] $. Our goal is to solve the Schr\"{o}dinger equation with singular
even-power and inverse-power potentials induced by the non-commutativity of
space. We thus find the exact non-commutative energy levels  and that the
non-commutativity effects are similar to the Zeeman splitting in commutative
space.

This paper is organized as follows. In section 2, we derive the deformed 2D
Schr\"{o}dinger equation for a central potentials $V\left( r\right)
=ar^{2}+br^{-2}+cr^{-4}$ and $V\left( r\right) =ar^{-1}+br^{-2}$ in
non-commutative space. We exactly solve the deformed Schr\"{o}dinger
equation in closed form \cite{39} and obtain the exact non-commutative
energy levels. Finally, section 3 is devoted to a discussion.

\section{Non-commutative Schr\"{o}dinger equation}

In this section we study the exact solutions of the Schr\"{o}dinger equation
for the potentials $V(r) =ar^2+br^{-2}+cr^{-4}$ and $V(r) =ar^{-1}+br^{-2}$
in the non-commutative space. The non-commutative model specified by eq. %
\eqref{eq:1} is defined by a star-product, where the normal product between
two functions is replaced by the $\star-$product: 
\begin{equation}
\left( \varphi \star \psi \right) \left( x\right) =\left. \varphi \left(
x\right) \exp\left(\frac{i}{2}\theta^{\mu \nu} \overleftarrow{\partial}_\mu 
\overrightarrow{\partial}_\nu\right) \psi \left( y\right) \right| _{x=y}.
\end{equation}

In a canonical non-commutative space-space type, the non-commutative quantum
mechanics is described by the following equation: 
\begin{equation}
H\left( p,x\right) \star \psi \left( x\right) =E\psi.
\end{equation}
This equation reduces to the usual one described by \cite{7,14}: 
\begin{equation}
H\left( \hat{p},\hat{x}\right) \psi \left( x\right) =E\psi,
\end{equation}
where 
\begin{equation}
\hat{x}_{i}=x_{i}-\frac{\theta _{ij}}{2}p_{j},\qquad\hat{p}_{i}=p_{i}.
\end{equation}

\subsection{ The potential $\boldsymbol{V( r) =ar^{2}+br^{-2}+cr^{-4}}$}

We can write the deformed potential $V\left( r\right) =ar^{2}+br^{-2}+cr^{-4}
$ in non-commutative space up to $\mathcal{O}\left( \Theta ^{2}\right) $ as: 
\begin{equation}
V\left( \hat{r}\right) =\frac{a}{2}\theta L_{z}+ar^{2}+br^{-2}+\left( c+%
\frac{b}{4}\theta L_{z}\right) r^{-4}+\frac{c}{2}\theta L_{z}r^{-6},
\end{equation}%
which is similar to the singular even-power potential which was studied in
ref \cite{39}.

The Schr\"{o}dinger equation in a 2D non-commutative space in the presence
of the potential $V\left( \hat{r}\right) $ can be cast into: 
\begin{equation}
\left( -\frac{1}{r}\frac{\partial }{\partial r}r\frac{\partial }{\partial r}-%
\frac{1}{r^{2}}\frac{\partial ^{2}}{\partial \varphi ^{2}}+V\left( \hat{r}%
\right) \right) \psi \left( \hat{r}\right) =E\psi \left( \hat{r}\right) .
\label{eq:KGmod}
\end{equation}%
The solution to eq. \eqref{eq:KGmod} in polar coordinates $(\hat{r},\hat{%
\varphi})$ takes the separable form \cite{39}: 
\begin{equation}
\psi \left( \hat{r}\right) =r^{-1/2}R_{m}\left( \hat{r}\right) e^{im\varphi
}.
\end{equation}%
Then eq. \eqref{eq:KGmod} reduces to the radial equation up to $\mathcal{O}%
\left( \Theta ^{2}\right) $ : 
\begin{equation}
\frac{d^{2}R_{m}\left( \hat{r}\right) }{dr^{2}}+\left[ \tilde{E}+\tilde{V}%
\left( r\right) -\frac{m^{2}-1/4}{r^{2}}\right] R_{m}\left( \hat{r}\right)
=0,  \label{eq:9}
\end{equation}%
where 
\begin{equation}
\tilde{V}\left( r\right) =ar^{2}+br^{-2}+\tilde{c}r^{-4}+\tilde{d}r^{-6},
\end{equation}%
and 
\begin{equation*}
\tilde{E}=E-\frac{a}{2}\theta m,\qquad \tilde{c}=c+\frac{b}{4}\theta
m,\qquad \text{and}\qquad \tilde{d}=\frac{c}{2}\theta m.
\end{equation*}

Equation \eqref{eq:9} is similar to the radial Schr\"{o}dinger equation with
singular even-power potential \cite{39}. To solve the equation %
\eqref{eq:KGmod}, we wite the radial functions as \cite{6,12}: 
\begin{equation}
R_{m}\left( \hat{r}\right) =e^{\tilde{p}_{\left\vert m\right\vert }\left(
r\right) }\sum_{n=0}\tilde{a}_{n}r^{2n+\tilde{\nu}},  \label{eq:11}
\end{equation}%
where 
\begin{equation}
\tilde{p}_{\left\vert m\right\vert }\left( r\right) =\frac{\alpha }{2}r^{2}+%
\frac{\tilde{\beta}}{2}r^{-2}.
\end{equation}%
Substituting eq. \eqref{eq:11} into eq. \eqref{eq:9} and equating the
coefficients of $r^{n+\nu }$ to zero, we obtain: 
\begin{equation}
\tilde{A}_{n}\tilde{a}_{n}+\tilde{B}_{n+1}\tilde{a}_{n+1}+\tilde{C}_{n+2}%
\tilde{a}_{n+2}=0,  \label{eq:13}
\end{equation}%
where 
\begin{eqnarray}
\tilde{A}_{n} &=&\tilde{E}+\alpha \left( 1+2\tilde{\nu}+4n\right)  \\
\tilde{B}_{n+1} &=&-b-2\alpha \tilde{\beta}-\left( m^{2}-\frac{1}{4}\right)
+\left( \tilde{\nu}+2n\right) \left( \tilde{\nu}-1+2n\right)  \\
\tilde{C}_{n} &=&\tilde{\beta}\left( 3-2\tilde{\nu}-4n\right) -\tilde{c},
\end{eqnarray}%
and 
\begin{equation}
\alpha ^{2}=a,\tilde{\beta}^{2}=\tilde{d}.
\end{equation}%
We can choose $\alpha $ and $\tilde{\beta}$ such that \cite{39}: 
\begin{equation}
\alpha =-\sqrt{a},\tilde{\beta}=\sqrt{|\tilde{d}|}.
\end{equation}

If $a_{0}\neq 0,$ then one obtains $C_{0}=0$, a condition that forbids the
existence of the $s$ energy levels ($\left\vert m\right\vert =2l+1$ in 2D).
This is in fact a particularity of the non-commutative Schr\"{o}dinger
equation solution, which is not present in the ordinary Schr\"{o}dinger
framework \cite{39}. Then we obtain: 
\begin{eqnarray}
\tilde{\nu} &=&\left( \frac{3}{2}+\frac{\tilde{c}}{2\tilde{\beta}}\right)  
\notag \\
&=&\left( \frac{3}{2}+\tilde{\gamma}\right) ,
\end{eqnarray}%
where 
\begin{equation}
\tilde{\gamma}=\frac{\tilde{c}}{2\tilde{\beta}}.
\end{equation}%
However if $a_{n}\neq 0,$ with $a_{n+1}=a_{n+2}=\cdots =0$ then $\tilde{A}%
_{n}=0$, from which one obtains the non-commutative energy eigenvalues exact
up to $\mathcal{O}\left( \Theta ^{2}\right) $ : 
\begin{equation}
\tilde{E}_{n,m}=\sqrt{a}\left( 4+2\tilde{\gamma}+4n\right) +\frac{a}{2}%
\theta m,\qquad \left\vert m\right\vert =1,2,3,\cdots .  \label{eq:21}
\end{equation}%
We have thus shown that the degeneracy with respect to the angular quantum
number $m$ is removed and that non-commutativity here acts like a Lamb shift.

Now, we discuss the corresponding exact solation for $n=1$. From eq. %
\eqref{eq:21} the non-commutative energy splitting of the energy levels up
to $\mathcal{O}\left( \Theta ^{2}\right) $ is: 
\begin{eqnarray}
\tilde{E}_{1,m} &=&\sqrt{a}\left( 8+\tilde{\gamma}\right) +\frac{a}{2}\theta
m  \notag \\
&=&\sqrt{a}\left( 8+\frac{c}{\sqrt{|\tilde{d}|}}\right) +\frac{a}{4}\left( 2+%
\frac{b}{\sqrt{a}}\right) \theta m  \notag \\
&=&\sqrt{a}\left( 8+\tilde{\lambda}\right) +\frac{a}{4}\left( 2+\delta
\right) \theta m,
\end{eqnarray}%
where 
\begin{equation}
\tilde{\lambda}=\frac{c}{\sqrt{|\tilde{d}|}},\qquad \delta =\frac{b}{\sqrt{a}%
}
\end{equation}

We have show that the non-commutative energy splitting is similar to the
Zeeman effects and removes the degeneracy with respect to $m$. Furthermore
we can say that the displacement of the energy levels is actually induced by
the space non-commutativity which plays the role of a magnetic field. The
corresponding eigenfunction is: 
\begin{equation}
\psi _{1}\left( \hat{r}\right) =\left( \tilde{a}_{0}+\tilde{a}
_{1}r^{2}\right) r^{\tilde{\nu}-1/2}e^{-\frac{1}{2}\left( \sqrt{a}r^{2}+%
\sqrt{| \tilde{d}| }r^{-2}\right) }e^{im\varphi },
\end{equation}
where $\tilde{a}_{0}$ and $\tilde{a}_{1}$ can be calculated from eq. %
\eqref{eq:13} and the normalisation condition. Following this method, we can
obtain a class of exact solutions. 

\subsection{The potential $\boldsymbol{V( r) =ar^{-1}+br^{-2}}$}

The deformed potential $V\left( r\right) =ar^{-1}+br^{-2}$ in
non-commutative space up to $\mathcal{O}\left( \Theta ^{2}\right) $ is: 
\begin{equation}
V\left( \hat{r}\right) =ar^{-1}+br^{-2}+\tilde{c}r^{-3}+\tilde{d}r^{-4},
\label{eq:25}
\end{equation}%
where 
\begin{equation}
\tilde{c}=\frac{a}{2}\theta m,\qquad \text{and}\qquad \tilde{d}=\frac{b}{4}%
\theta m,
\end{equation}%
where the third term is the dip\^{o}le-dip\^{o}le interaction created by the
non-commutativity, the second term is a similar to the interaction between
an ion and a neutral atom created by the non-commutativity. These
interactions show us that the effect of space non-commutativity on the
interaction of a single-electron atom, for example, is similar to that of a
charged ion interacting with the atom on the one hand and on the other hand
interacting with the electron to create a dipole and with the nucleus to
create a second dipole. 

The approach of the potential in eq. \eqref{eq:25} is similar to that for
the inverse-power potential in a commutative space. Thus we can take as
solutions the eigenfunctions from Ref. \cite{40}: 
\begin{equation}
R_{m}\left( \hat{r}\right) =h_{m}\left( \hat{r}\right) e^{f\left( \hat{r}%
\right) },\,m=1,2,3,\cdots 
\end{equation}%
where 
\begin{equation}
f\left( \hat{r}\right) =Ar^{-1}+Br+C\log r,\,A\prec 0\quad \text{and}\quad
B\prec 0,
\end{equation}%
and 
\begin{equation}
h_{m}\left( \hat{r}\right) =\prod\limits_{j=1}^{m}\left( r-\tilde{\sigma}%
_{j}^{m}\right) =\sum_{j=1}^{m}\tilde{a}_{j}r^{j}.
\end{equation}

Then the radial Schr\"{o}dinger eq. \eqref{eq:21} reduces to the following
equation: 
\begin{equation}
\left[ f^{^{\prime \prime }}+f^{^{\prime }2}+\frac{h_{m}^{^{\prime \prime
}}+2h_{m}^{^{\prime }}f^{^{\prime }}}{h_{m}}+E-V\left( \hat{r}\right) -\frac{%
m^{2}-1/4}{r^{2}}\right] R_{m}\left( \hat{r}\right) =0.
\end{equation}%
We arrive at the equation \cite{40}: 
\begin{equation}
f^{^{\prime \prime }}+f^{^{\prime }2}+\frac{h_{m}^{^{\prime \prime
}}+2h_{m}^{^{\prime }}f^{^{\prime }}}{h_{m}}=-E+V\left( \hat{r}\right) -%
\frac{m^{2}-1/4}{r^{2}}.  \label{eq:31}
\end{equation}%
Now using the fact that: 
\begin{equation}
f^{^{\prime \prime }}+f^{^{\prime }2}=B^{2}+\frac{2BC}{r}-\frac{2AB}{r^{2}}+%
\frac{2A-2AC}{r^{3}}+\frac{A^{2}}{r^{4}},  \label{eq:32}
\end{equation}%
and 
\begin{equation}
h_{m}^{^{\prime }}=\sum_{j=1}^{m}j\tilde{a}_{j}r^{j-1},\text{ \ \ \ }%
h_{m}^{^{\prime \prime }}=\sum_{j=1}^{m}j\left( j-1\right) \tilde{a}%
_{j}r^{j-2},  \label{eq:33}
\end{equation}%
where 
\begin{eqnarray}
a_{m} &=&1  \notag \\
a_{m-1} &=&-\sum_{j=1}^{m}\tilde{\sigma}_{j}^{m}  \label{eq:34} \\
a_{m-2} &=&-\sum_{j\prec i}^{m}\tilde{\sigma}_{j}^{m}\tilde{\sigma}_{i}^{m} 
\notag
\end{eqnarray}%
and so on, then eqs. \eqref{eq:31}-\eqref{eq:33} lead to an algebraic
equation where we equate equivalent coefficients of $r^{s}$ between both
sides of the equation, taking into account the eq. \eqref{eq:34}, we find: 
\begin{align}
& A^{2}=\tilde{d},\qquad \tilde{E}=-B^{2}  \label{eq:35} \\
& 2A\left( 1-C\right) =\tilde{c} \\
& a=2B\left( C+m\right) ,  \label{eq:37}
\end{align}%
and 
\begin{equation}
\lambda =b+m^{2}-1/4=C\left( C+2m-1\right) +m\left( m-1\right) -2B\left(
A-\sum_{j=1}^{m}\sigma _{j}^{m}\right) ,  \label{eq:38}
\end{equation}%
and 
\begin{equation}
m\sqrt{\tilde{d}}+\left( m+1+C\right) \sum_{j=1}^{m}\sigma
_{j}^{m}+B\sum_{j=1}^{m}\left( \sigma _{j}^{m}\right) ^{2}=0,
\end{equation}%
\begin{equation}
\left( m-1\right) \sqrt{\tilde{d}}\sum_{j=1}^{m}\sigma _{j}^{m}+2\left(
m-1+C\right) \sum_{j\prec i}^{m}\sigma _{j}^{m}\sigma _{i}^{m}+B\sum_{j\prec
i}^{m}\sigma _{j}^{m}\sigma _{i}^{m}\sum_{l\prec k}^{m}\left( \sigma
_{l}^{m}+\sigma _{k}^{m}\right) =0,
\end{equation}%
\begin{multline}
\left( m-2\right) \sqrt{\tilde{d}}\sum_{j\prec i}^{m}\sigma _{j}^{m}\sigma
_{i}^{m}+3\left( m-2+C\right) \sum_{j\prec i\prec k}^{m}\sigma
_{j}^{m}\sigma _{i}^{m}\sigma _{k}^{m}+ \\
+B\sum_{j\prec i\prec k}^{m}\sigma _{j}^{m}\sigma _{i}^{m}\sigma
_{k}^{m}\sum_{l\prec q\prec s}^{m}\left( \sigma _{l}^{m}+\sigma
_{q}^{m}+\sigma _{s}^{m}\right) =0,
\end{multline}

Moreover, multiplying equation \eqref{eq:38} by $B$ and using eqs.%
\eqref{eq:35} - \eqref{eq:37} we find the following algebraic equation for $B
$ as: 
\begin{equation}
4\left( A-\sum_{j=1}^{m}\sigma _{j}^{m}\right) B^{2}+2\tilde{\omega}%
B-a\left( \tilde{\nu}+2m\right) =0,  \label{eq:42}
\end{equation}%
where 
\begin{equation}
\tilde{\omega}=\lambda +m\left( 2m+\tilde{\nu}\right) -m\left( m-1\right) ,
\end{equation}%
and 
\begin{equation}
\tilde{\nu}=C-1=\frac{\tilde{c}}{2\sqrt{\tilde{d}}}.
\end{equation}%
The equation \eqref{eq:42} is solved by: 
\begin{eqnarray}
B_{\pm } &=&\frac{\tilde{\omega}\pm \sqrt{\tilde{\omega}^{2}+4\left(
A-\sum_{j=1}^{m}\sigma _{j}^{m}\right) a\left( \tilde{\nu}+2m\right) }}{%
4\left( A-\sum_{j=1}^{m}\sigma _{j}^{m}\right) }  \notag \\
&=&\tilde{\omega}\frac{1\pm \sqrt{1+4\left( A-\sum_{j=1}^{m}\sigma
_{j}^{m}\right) \frac{a\left( \tilde{\nu}+2m\right) }{\tilde{\omega}^{2}}}}{%
4\left( A-\sum_{j=1}^{m}\sigma _{j}^{m}\right) }.
\end{eqnarray}%
So the non-commutative energy spectrum up to $\mathcal{O}\left( \Theta
^{2}\right) $  is given by: 
\begin{equation}
\tilde{E}=-\tilde{\omega}^{2}\frac{\left( 1\pm \sqrt{1+4\left(
A-\sum_{j=1}^{m}\sigma _{j}^{m}\right) \frac{a\left( \tilde{\nu}+2m\right) }{%
\tilde{\omega}^{2}}}\right) ^{2}}{16\left( A-\sum_{j=1}^{m}\sigma
_{j}^{m}\right) ^{2}},
\end{equation}%
where 
\begin{equation}
\tilde{\omega}^{2}=\theta m\frac{a^{2}}{4b}\left( m^{2}+\frac{2m}{\tilde{\nu}%
}\left( \lambda +m\left( m+1\right) \right) \right) +\left( \lambda +m\left(
m+1\right) \right) ^{2}.
\end{equation}

We have thus shown that the non-commutativity effects are manifested in
energy levels, so that they are split into $m$ levels, similarly to the
effects of the magnetic field. Thus we can say that the non-commutativity
plays the role of the magnetic field. It is also found that if the limit $%
\theta \rightarrow 0$ is taken, then we recover the results of the
commutative case \cite{40}.

\section{Conclusions}

In this paper we started from a quantum particle with the central potentials 
$V\left( r\right) =ar^{2}+br^{-2}+cr^{-4}$ and $V\left( r\right)
=ar^{-1}+br^{-2}$ in a canonical non-commutative space. Using the Moyal
product method, we have derived the deformed Schr\"{o}dinger equation, we
showed that it a similar to the Schr\"{o}dinger equation with singular
even-power and inverse-power potentials in commutative space. Using the
power-series expansion method we solved it exactly and we found that the
non-commutative energy is shifted to $m$ levels. The non-commutativity acts
here like a Lamb shift. This proofs that the non-commutativity has an effect
similar to the Zeeman effects, where the non-commutativity leads the role of
the magnetic field.


\begin{thebibliography}{99}
\bibitem{1} M Chaichian, M M Sheikh-Jabbari and A Tureanu \textit{Phys. Rev.
Lett}. \textbf{86} 2716 (2001).

\bibitem{2} T C Adorno et al. \textit{Phys. Lett. B} \textbf{682} 235 (2009).

\bibitem{3} H Motavalli, A R Akbarieh \textit{Mod. Phys. Lett. A} \textbf{25}
2523 (2010).

\bibitem{3a} R. Eid, S. I. Muslih, D. Baleanu, Romanian Journal of Physics
56 (3--4), 323--331 (2011).

\bibitem{3b} S. Al-Jaber, Rom. Journ. Phys., Vol. 58, Nos. 3--4, P. 247--259
(2013).

\bibitem{4} J. Audretsch and R. M"uller Phys. Rev. A52, 629 (1995).

\bibitem{5} R. Passante Phys. Rev. A\textbf{57}, 1590 (1998).

\bibitem{6} L. Rizzuto Phys. Rev. A\textbf{76}, 062114 (2007).

\bibitem{7} Z. Zhu and H. Yu Phys. Rev. A\textbf{82}, 042108 (2010).

\bibitem{8} A. Share and S. N. Behra, Pramana J. Phys. 14 (1980).

\bibitem{9} D. Amin, Phys. Today 35, 35(1982); Phys. Rev. Lett. 36, 323
(1976).

\bibitem{10} S. Coleman, \textquotedblleft Aspects of
Symmetry\textquotedblright\ selected Erice Lectures (Cambridge Univ.Press,
Cambridge, 1988), p. 234.

\bibitem{11} H. Hashimoto, Int. J. Electron 46, 125 (1979), Opt. Commu. 32,
383 (1980).

\bibitem{12} C. E. Reid, J. Mole. Spectro. 36, 183 (1970).

\bibitem{13} R. S. Kaushal, Ann. Phys. (N. Y. ) 206, 90 (1991).

\bibitem{14} R. S. Kaushal, and D. Parashar, Phys. Lett. A 170, 335 (1992).

\bibitem{15} R. S. Kaushal, Phys. Lett. A 142, 57 (1989).

\bibitem{16} S. K. Bose and N. Varma, Phys. Lett. A 147, 85 (1990).

\bibitem{17} S. K. Bose, IL Nuovo Cimento B 109, 1217 (1994).

\bibitem{18} A. Voros, J. Phys. A. 32, 5993 (1999).

\bibitem{19} Y. P. Varshni, Phys. Lett. A 183, 9 (1993).

\bibitem{20} S. 
\"{}%
Ozcelik and M. Simsek, Phys. Lett. A 152, 145 (1991).

\bibitem{21} M. Simsek and S. 
\"{}%
Ozcelik, Phys. Lett. A 186, 35 (1994).

\bibitem{22} M. Simsek, Phys. Lett. A 259, 215 (1999).

\bibitem{23} Shi-Hai Dong and Zhong-Qi Ma, J. Phys. A. 31, 9855 (1998).

\bibitem{24} Shi-Hai Dong, Zhong-Qi Ma and G. Esposito, Found. Phys. Lett.
12, 465 (1999).

\bibitem{25} M. Znojil, J. Math. Phys. 30, 23 (1989).

\bibitem{26} M. Znojil, J. Math. Phys. 31, 108 (1990).

\bibitem{27} M. Znojil, J. Phys. A 15, 2111 (1982).

\bibitem{28} V. de Alfaro and T. Regge, Potential Scattering (North Holland,
Amsterdam,1965).

\bibitem{29} S. Fubini and R. Stroffolini, Nuovo Cimento 37, 1812 (1965).

\bibitem{30} F. Calogero, Variable phase Approach to Potential Scattering
(Academic, N. Y.,1967).

\bibitem{31} R. G. Newton, Scattering Theory of Waves and Particles (McGraw
Hill, N. Y.,1967).

\bibitem{32} W. M. Frank, D. J. Land and R. M. Spector, Rev. Mod. Phys. 43,
36 (1971).

\bibitem{33} R. Stroffolini, Nuovo Cimento A 2, 793 (1991).

\bibitem{34} G. Esposito, J. Phys. A 31, 9493 (1998).

\bibitem{35} G. Esposito, Found. Phys. Lett. 11, 535 (1998).

\bibitem{36} A. O. Barut, J. Math. Phys. 21, 568 (1980).

\bibitem{37} B. H. Bransden and C. J. Joachain, Physics of Atoms and
Molecules (Longman,London, 1983).

\bibitem{38} G. C. Maitland, M. M. Higby, E. B. Smith and V. A. Wakoham,
Intermolecular Forces (Oxford University Press, Oxford, 1987).

\bibitem{39} Shi-Hai Dong,I.J.Theo. Phys,Vol. 39,No,4, (200).

\bibitem{40} Shi-Hai Dong, Zhong-Qi Ma, Giampiero Esposito, Phys.Lett. 12
(1999) 465-474.
\end{thebibliography}
\end{document}